\documentclass{article}

\usepackage{PRIMEarxiv}
\usepackage[utf8]{inputenc} 
\usepackage[T1]{fontenc}    
\usepackage{hyperref}       
\usepackage{url}            
\usepackage{booktabs,caption}       
\usepackage{amsfonts}       
\usepackage{nicefrac}       
\usepackage{microtype}      
\usepackage{lipsum}
\usepackage{fancyhdr}       
\usepackage{graphicx}       
\usepackage{array}
\usepackage{easyReview}
\usepackage{multirow}
\usepackage{tablefootnote}
\usepackage{pgfplots}
\usepackage{amsmath} 
\pgfplotsset{compat=1.3}
\usepackage[flushleft]{threeparttable}
\usepackage{newunicodechar}
\graphicspath{{media/}}     
\newtheorem{remark}{Remark}

\newcommand{\eg}{\mbox{{\em e.g.}}}

\newcommand{\proc}[1]{\ifmmode\mbox{\textsc{#1}}\else\textsc{#1}\fi}
\newcommand{\sentry}{\proc{CITADEL}}
\title{Security Enclave Architecture for Heterogeneous Security Primitives for Supply-Chain Attacks
}

\author{
  Kshitij Raj\\
  University of Florida\\
  Gainesville, FL, USA\\
  \texttt{kshitijraj@ufl.edu} \\
   \And
  Atri Chatterjee\\
  University of Florida\\
  Gainesville, FL, USA\\
  \texttt{a.chatterjee@ufl.edu} \\
     \And
  Patanjali SLPSK\\
  University of Florida\\
  Gainesville, FL, USA\\
  \texttt{atanjal.sristil@ufl.edu} \\
     \And
  Swarup Bhunia\\
  University of Florida\\
  Gainesville, FL, USA\\
  \texttt{swarup@ece.ufl.edu} \\
     \And
  Sandip Ray\\
  University of Florida\\
  Gainesville, FL, USA\\
  \texttt{sandip@ece.ufl.edu} \\
}

\begin{document}
\maketitle

\begin{abstract}
Designing secure architectures for system-on-chip (SoC) platforms is a highly intricate and time-intensive task, often requiring months of development and meticulous verification. Even minor architectural oversights can lead to critical vulnerabilities that undermine the security of the entire chip. In response to this challenge, we introduce **CITADEL**, a modular security framework aimed at streamlining the creation of robust security architectures for SoCs. CITADEL offers a configurable, plug-and-play subsystem composed of custom intellectual property (IP) blocks, enabling the construction of diverse security mechanisms tailored to specific threats. As a concrete demonstration, we instantiate CITADEL to defend against supply-chain threats, illustrating how the framework adapts to one of the most pressing concerns in hardware security. This paper explores the range of obstacles encountered when building a unified security architecture capable of addressing multiple attack vectors and presents CITADEL’s strategies for overcoming them. Through several real-world case studies, we showcase the practical implementation of CITADEL and present a thorough evaluation of its impact on silicon area and power consumption across various ASIC technologies. Results indicate that CITADEL introduces only minimal resource overhead, making it a practical solution for enhancing SoC security.
\end{abstract}


\section{Introduction}
\label{introduction}
With increasing globalization of the design and fabrication processes, the development of a modern microelectronic involves a complex supply chain with a large number of participants dispersed geographically across the world.  Most computing devices are developed through System-on-Chip (SoC) designs, which entail the integration of pre-designed hardware blocks (referred to as ``intellectual properties'' or ``IPs'') into a single silicon substrate.  Players in the SoC design supply chain include  IP vendors and suppliers, SoC integration houses, foundries, and testing facilities,  Original Equipment Manufacturers (OEMs), and product suppliers, among others.  There can be rogue entities in this complex ecosystem that can attempt to subvert the security of the SoC product or operations of other players in the ecosystem.  Such subversions can include implanting malicious circuitry or Trojans in the design, reverse-engineering the design functionality, fabricating counterfeit designs, and many others.  Obviously, it is crucial to develop security architectures to protect against such attacks.

\begin{figure*}[h]
		\begin{center}
			\centering
			\includegraphics[width=0.9\textwidth]{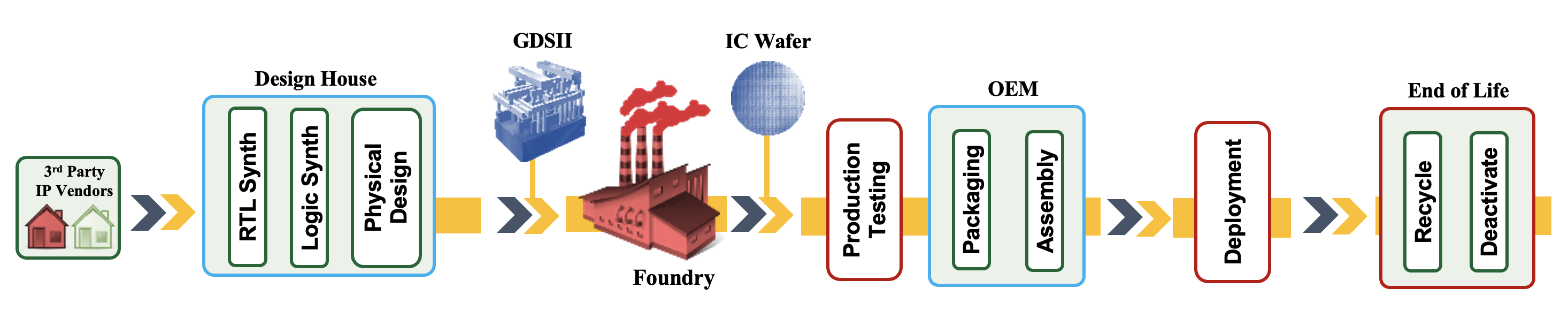}
		\end{center}
		\caption{SoC Lifecycle from Supply Chain Perspective. Only major players in the supply chain are indicated.}
		\label{fig:lifecycle}
\end{figure*} 

There has been significant research over the past decade to detect and mitigate supply-chain attacks, resulting in numerous interesting approaches.  This includes design obfuscation algorithms, various IC and IP authentication techniques, design watermarking, trojan detection, and many others. Correspondingly, numerous security engines have also been developed that produce secure enclaves targeting specific threat vectors, {\itshape e.g.} AEGIS \cite{AEGIS}, Sanctum \cite{sanctum} and Komodo \cite{komodo} provide isolation for trusted execution environments (TEEs), primarily revolving around secure software execution in an untrusted environment. However, --- and in spite of its critical need, --- we have not found (parameterized) security architectures for systematically integrating heterogeneous solutions into SoC designs or enforcing them at different stages in the design lifecycle for different kinds of security solutions, and no specific architecture targeting an untrusted supply-chain (See Section \ref{related-work}).

In this paper, we develop an architectural framework for the disciplined incorporation of security requirements in modern SoC designs.  The key idea is to develop a {\itshape skeletal architecture} that can be configured to a variety of security use cases through the integration of custom {\itshape security countermeasures (SCMs)}.  We exploit this framework to create a systematic, plug-and-play, and configurable architecture, called ``$\sentry$'' for mitigating supply chain threats across the SoC lifecycle.  Furthermore, our experimental results demonstrate that $\sentry$ incurs a very small hardware footprint, making it viable for low-power SoC designs.

The paper makes the following important contributions.
\begin{itemize}
    \item $\sentry$ represents to our knowledge the first comprehensive architecture that enables the integration of ad-hoc supply chain security protection in SoC designs through a configurable plug-and-play subsystem. Furthermore, our work represents the first parameterized architectural framework for the systematic integration of SoC security primitives.
    \item We develop a comprehensive threat model for SoC designs that accounts for different lifecycles, and show how the $\sentry$ solution can enable systematic protection from the spectrum of threats at each lifecycle.
    \item We demonstrate the viability of the $\sentry$ solution through several security case studies as well as detailed overhead analysis on multiple ASIC implementations.
\end{itemize}

The rest of the paper is organized as follows. Section \ref{background} provides the relevant background on supply chain threats and challenges across the SoC lifecycle. Section \ref{skeletal} gives a high-level overview of our parameterized skeletal architecture and the operating ecosystem. Section \ref{sentrysc} describes the $\sentry$ solution in detail and explains how it incorporates systematic integration of SoC security functionality to mitigate supply-chain threats. In Section \ref{usecase}, we provide detailed use cases demonstrating the versatility of  $\sentry$. Section \ref{experiment} provides our experimental results for the evaluation of $\sentry$ overhead. We discuss related work in Section \ref{related-work} and conclude in Section \ref{conclusion}.

\section{Background}
\label{background}
Over the past years, microelectronics design has evolved into a globally distributed supply chain incorporating 3PIP (third-party IP) vendors, IC design houses, fabrication plants, and assembly, packaging, and testing facilities. Fig. \ref{fig:lifecycle} traces the lifecycle of a microelectronics system across the supply chain.  While the globalization of the supply chain has been instrumental in ameliorating the challenges arising from increased design complexities, aggressive time-to-market schedules, etc., an unfortunate side-effect is the growing problem of assurance across the complex supply chain  In this section, we briefly recount some categories of supply-chain threats relevant to the work in this paper.  Table \ref{tab:threat-handled} provides a summary of the threats involved, followed by a brief description.

\begin{table}[]
\centering
\caption{Threat Classification, Mitigation Strategies, and Asset Categories used by $\sentry$ for Supply-Chain}
\label{tab:threat-handled}
\def\arraystretch{1.2}
\begin{tabular}{llll}
\hline
\textbf{Class of Threat} & \textbf{Potential Adversaries} & \textbf{Mitigation Approach} & \textbf{Mitigation Technique} \\ \hline
\begin{tabular}[c]{@{}l@{}}Counterfeiting \\ (Piracy and Cloning), \\ Overproduction\end{tabular} &
  Fabrication Facility &
  \begin{tabular}[c]{@{}l@{}}Unique and Unclonable \\ Chip Identifier\end{tabular} &
  \begin{tabular}[c]{@{}l@{}}Unique\& Unclonable \\ ChipID obtained using \\ embedded MeLPUF\end{tabular} \\ \hline
Reverse Engineering &
  \begin{tabular}[c]{@{}l@{}}Testing Facility, \\ Recycling Facility\end{tabular} &
  \begin{tabular}[c]{@{}l@{}}Design Obfuscation, \\ Redaction\end{tabular} &
  \begin{tabular}[c]{@{}l@{}}ProtectIP State-Space \\ Obfuscation\end{tabular} \\ \hline
Illegal Recycling &
  \begin{tabular}[c]{@{}l@{}}Testing Facility, \\ Recycling Facility\end{tabular} &
  \begin{tabular}[c]{@{}l@{}}Chip Lifecycle Identification\\  and Tracking\end{tabular} &
  \begin{tabular}[c]{@{}l@{}}Lifecycle validation \& \\ tracking with AMI\end{tabular} \\ \hline
\end{tabular}
\end{table}

\subsection{Threat Classification}
\subsubsection{Overproduction and Counterfeiting}  An untrusted foundry with access to the design information (layout) may produce counterfeit ICs, including recycled, remarked, and cloned chips that are then sold as legitimate ones to the end user.  Note that these overproduced chips, which are then illegally sold in the market, have the same specifications as the original, but the foundry does not have an obligation to meet the functional requirements and carry out extensive validation, as these are sold in the black market or by other interested parties.

\subsubsection{Reverse Engineering} Foundries have control over the manufacturing process and can reverse engineer the chips to steal IP design secrets, expose assets, and derive implementation details by deconstructing the die layer by layer. The severity of reverse engineering depends on the lifecycle of the chip. The netlist can be extracted directly from the layout information (GDSII file) at the foundry \cite{rajarathnam2020regds}, enabling malicious parties to implant Trojans or alter the functionality of the chip. A malicious testing facility may reverse engineer the chip to steal Chip IDs, cryptographic assets, etc. The situation is further exacerbated by modern-day synthesis techniques, which often leave behind the execution trace of circuit functionality, enabling malicious parties to backtrack the synthesis and steal design specifics.

\subsubsection{Illegal Recycling} The global supply chain has seen a sharp rise in the practice and number of recycled chips being re-branded and re-packaged as new and sold, posing a serious threat to commercial and military applications. The harvested chips have inferior quality, security assurance, shorter lifespan, and performance degradation over their lifespan. A 2020 survey done by the Government \& Industry Data Exchange Program found a 154\%  increase in the number of counterfeit ICs traded internationally and a 10-fold increase in seizures of such counterfeit ICs at U.S. borders \cite{dhsreport}.

\subsection{A PUF Implementation}
\begin{figure}
	\begin{center}
		\centering
		\includegraphics[width=0.6\columnwidth]{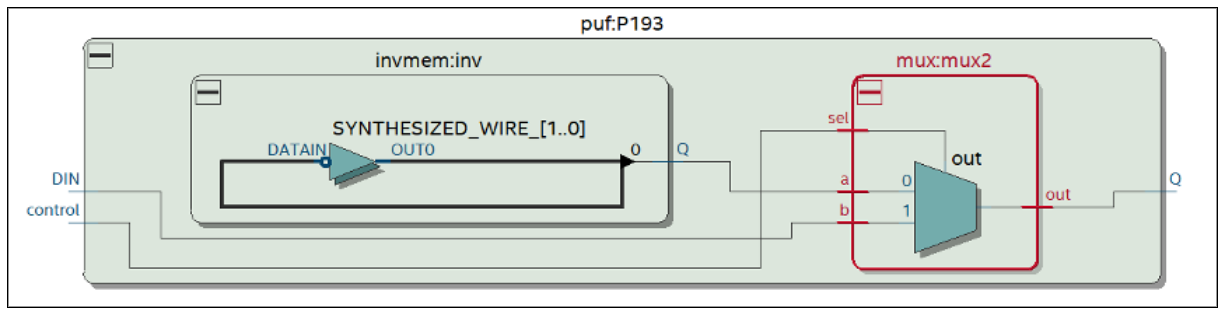}
	\end{center}
	\caption{RTL schematic of the MeLPUF architecture generated in Intel Quartus\textsuperscript{\textregistered} 18.1. The bi-stable memory element connects to the input of the 2x1 MUX with the circuit data input connected in b. The 2x1 MUX acts as the control element with the control signal being the input to select.}
	\label{fig:pufcircuit}
\end{figure}

Although the concepts behind the security architecture presented in this paper are general, we focus on instantiating it with a specific architectural instance intended to protect SoC designs against supply-chain attacks (see Section \ref{sentrysc}). A key security IP used in that instantiation is a PUF module, which creates a unique, unclonable chip ID.  The specific PUF module used in our evaluation and experiments is MeLPUF \cite{melpuf}, shown in Fig. \ref{fig:pufcircuit}.  The PUF exploits the observation that every chip die has a unique doping density and these variations at the sub-micron level result in different capacitance values in the uninitialized bi-stable memory cells. By integrating the cells in specific IPs of an SoC, MeLPUF can generate a unique, unclonable ID for the IP which can act as a watermark; furthermore, by distributing MeLPUF cells across multiple IPs in the SoC, it is possible to collect a signature of the entire SoC that can act as a ChipID. Each MeLPUF bit incurs an overhead of 6 NAND gates.

\begin{figure}
\centering
\begin{minipage}{.5\textwidth}
  \centering
  \includegraphics[width=\columnwidth]{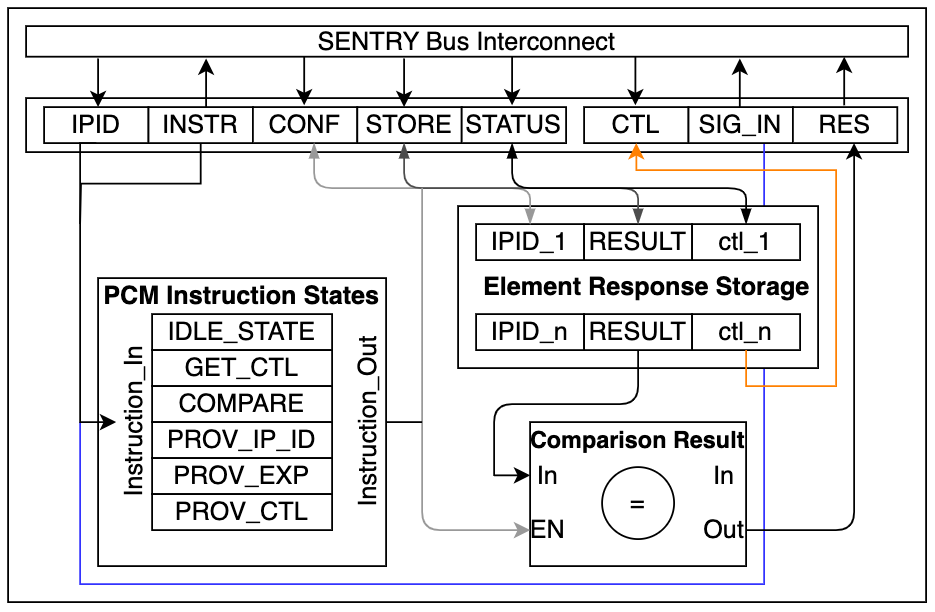}
  \caption{An Overview of the circuitry for the PUF\\ Control Module (PCM).}
  \label{fig:pcmcircuit}
\end{minipage}%
\begin{minipage}{.5\textwidth}
  \centering
  \includegraphics[width=\columnwidth]{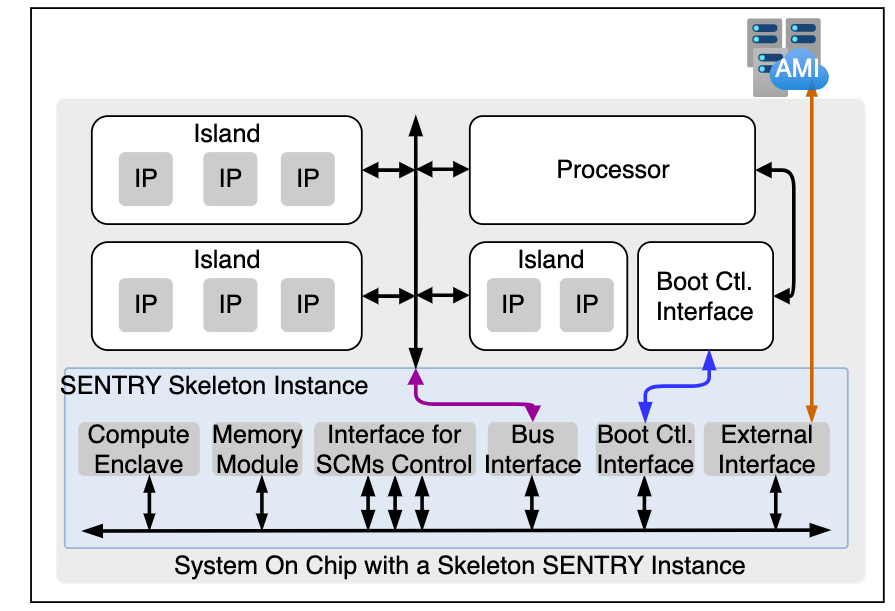}
  \caption{The $\sentry$ skeletal architecture integrated with a representative SoC model.}
  \label{fig:sentry_skeleton}
\end{minipage}
\end{figure}

Realizing such a PUF in an SoC requires, in addition to the distributed MeLPUF cells, a PUF Control Module (PCM) for sending, receiving, storing, and authenticating the PUF challenge-response pairs.  Fig.~\ref{fig:pcmcircuit} shows a representative implementation of PCM.  During the enrollment phase, the PCM provisions the IP IDs, control signals, and the expected responses. IP IDs are first provisioned and stored in the PCM buffer and are in a 1-1 correspondence with all designated IPs. These IDs are then used to index the control signals and expected responses according to their corresponding IP IDs. The PCM transmits the control signal and collects the PUF responses from the IPs. It then compares the signatures with the results of the authentication operation. The PCM also incorporates an error correction module. Error correction is performed on the incoming PUF signatures during authentication. The parity bits for error correction are calculated from the expected signature. The error correction module takes 16 bits at a time, therefore, the PUF signature is split into 16-bit segments for correction. From an architectural perspective, observe that the implementation of the PUF above requires a strategy for integrating individual MeLPUF cells with IPs in the SoC while the PCM can be designed as a standalone IP to perform the coordination and error correction functionality.  We exploit this observation in Section \ref{sentrysc} by integrating MeLPUF cells with our security wrapper architecture and using the PCM as a standalone IP.

\subsection{Design Obfuscation Basics}
Hardware Design obfuscation is a technique to protect design IPs from reverse-engineering attacks.  Here, an IP is instrumented so that the desired functionality is only realized when a specific input sequence is first applied. The input sequence is often referred to as a {key} and the application of the key as an initial input sequence is called {unlocking}.  The idea is that the key is not known to the malicious or untrusted party, (foundry), and consequently, the desired functionality is protected from reverse-engineering. Multiple obfuscation schemes have been proposed in two broad categories: (i) combinational obfuscation, where the combinational logic of the design is obfuscated \cite{protectip}, and (ii) sequential obfuscation, where the same applies to the sequential components of the design \cite{statelock}.
\section{The $\sentry$ Skeletal Architecture}
\label{skeletal}

The $\sentry$ architecture is based on a {skeletal architectural framework} which can be instantiated and configured through individual security countermeasure (SCM) IPs to realize a variety of protection functionalities.  The key observation is that the realization of (any) SoC security architecture entails the following ingredients:
\begin{itemize}
    \item A centralized control unit.
    \item An interface for secure communication with HOST IPs.
    \item An interface for secure off-chip communication.
\end{itemize}

Each of the ingredients above in fact has specific essential and intuitive functionality.  Implementing and enforcing security countermeasures requires a control unit for coordinating and enforcing security countermeasures, hand-offs, data movement across SCMs, and interactions among various entities in the SoC.  Any countermeasure involves coordination and communication of the control unit with different IPs, requiring a standardized communication interface.  Finally, checking the authenticity of the chip or any hardware IP in the SoC at any point during its operation requires communication with a trusted off-chip entity that keeps track of the transitions of the chip at different lifecycles. 

\begin{figure}
		\begin{center}
			\centering
			\includegraphics[width=0.7\columnwidth]{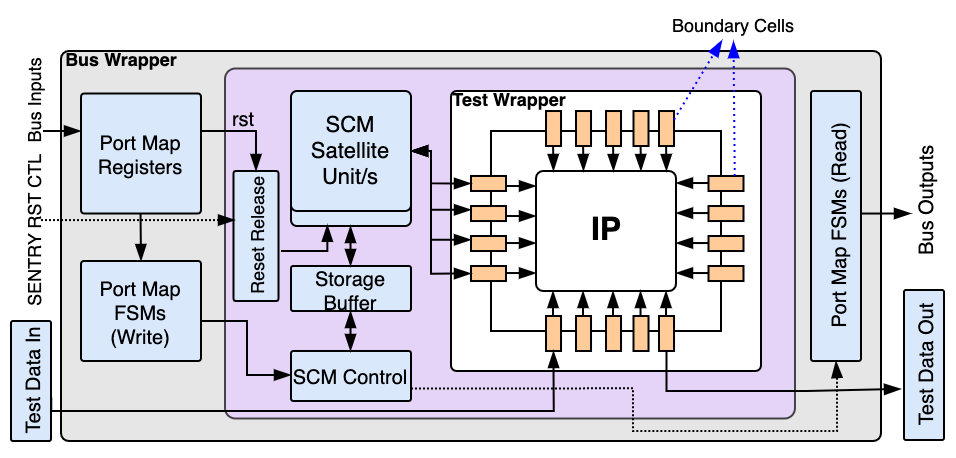}
		\end{center}
		\caption{A generic security wrapper with SCM enforcement support.}
		\label{fig:wrapper_skeleton}
\end{figure}

$\sentry$ incorporates this observation through the skeletal architecture shown in Fig.~\ref{fig:sentry_skeleton}.  The control unit is realized through a Compute Enclave (CE) that coordinates and orchestrates actions involved in enforcing security countermeasures.  The enclave is implemented with a (lightweight) microprocessor with a standardized interconnect to connect to various SCMs (see below), and its operation and control can be configured through security policies \cite{yier} implemented by firmware. The communication of the compute enclave with the IPs in the SoC is implemented through a standardized {wrapper}, as shown in Fig. \ref{fig:wrapper_skeleton}. The key intuition is that the realization of (any) wrapper to enable communication entails the following:

\begin{itemize}
    \item An parameterized interface to communicate with the external entity.
    \item A mechanism to communicate with the internal entity (IP).
    \item In case of security enforcement, a mechanism to carry out security operations to the underlying IP. This may include accessing specific registers, flip-flops, and other data elements in the IP to extract and/or program specific values at various execution phases. 
\end{itemize}

This generic wrapper enables the enclave to communicate with IPs independent of the underlying IP implementation through dedicated \texttt{PORT MAP READ/WRITE} FSMs, which can be customized based on the interface implementation. To facilitate interaction with the underlying IP, the wrappers in $\sentry$ are implemented as an extension for the IEEE-1500 test wrapper already available with most IPs. The $\sentry$ compute enclave interacts with the IPs through these communication interfaces (see below). Any transaction $t$ from $\sentry$ to another IP $P$ in the SoC is interpreted by the wrapper in $P$, which translates $t$ to operations that need to be performed by $P$. 

While these wrappers are generic, they can be augmented to provide support for security enforcement by integrating SCM satellite units and control, as shown in Fig. \ref{fig:wrapper_skeleton}, to realize security wrappers. The key realization here is that SCMs (may) have a distributed implementation mechanism, which although controlled by a centralized unit, has elements spread across IPs in the SoC. Activating and controlling those elements may require access to specific flip-flops or registers of the IP. The wrappers leverage the fundamental idea that test wrappers have access to low-level registers and flip-flops of design elements at specific instances during execution, using which SCM control and data extraction can be facilitated. This enables control of distributed SCM endpoints by a centralized control unit ($\sentry$). The SCMs in the wrapper require two distinct interfaces for communication with $\sentry$ and the test wrapper to extract or control data from the underlying IP (shown in Fig. \ref{fig:wrapper_skeleton}). As such, they are sandwiched between a bus wrapper, which facilitates communication with $\sentry$ to receive control packets, and the test wrapper, which performs data extraction and application to the underlying IP for SCM enforcement. The generic wrapper also includes a storage buffer and an SCM control unit to coordinate and manage distributed SCM unit/s in the wrapper, which can be configured through fine-grained security policies. The generic wrapper incorporates a reset release switch, enabling $\sentry$ to hold IPs at reset (using the \texttt{SENTRY RST CTL} signal) even when the HOST processor has released the reset signal. Reset gating enables $\sentry$ to configure and control IPs at boot time to control boot-critical events.

\smallskip

\paragraph{Interface for Secure Off-chip Communication} The $\sentry$ ecosystem includes a cloud-based {Asset Management Infrastructure} (AMI) for storing and retrieving assets and chip metadata (ChipIDs, obfuscation keys, cryptographic keys, authorization status). $\sentry$ supports secure communication with AMI through an Ethernet interface.

\begin{remark}
    One view of the architecture is that the $\sentry$ subsystem acts as a mediator for communication of the IPs with the external world, \eg, AMI.  Obviously, an alternate architectural strategy could be for a communication channel to be directly established between IPs (using the IEEE 1500 wrappers) and the AMI. Such an architecture would get away without using $\sentry$ to mediate over this communication. However, a deeper analysis reveals that the communication between the IPs and the AMI needs to account for a {\em controlled and coordinated} information flow mechanism. Doing so without having a mediator (in this case $\sentry$) would lead to a complex distributed architecture and require humongous efforts to coordinate any such communication to avoid potential security challenges. In addition to this, the overheads of such an architecture would be significantly more than $\sentry$.
\end{remark}
\smallskip

The skeletal architecture discussion above might seem simplistic at first glance.  However, there are interesting subtleties that must be addressed to ensure the architecture is indeed viable.  An interesting conundrum is the design of the communication interfaces between $\sentry$ and the HOST SoC.  Note that the role of $\sentry$ is to be in control only of the security functionality of the SoC; other functionalities such as power management or sleep/wakeup needs of the different IPs and communication fabrics are beyond $\sentry$'s purview\footnote{$\sentry$ compute enclave is in control of the power management of the $\sentry$ subsystem, including the internal bus.  However, it cannot control the power management of communication fabrics outside the subsystem.}.  However, the ability of $\sentry$ to communicate with an IP through the system bus is constrained by whether the bus is active or asleep at a certain point in the system execution.  A key challenge is the system boot, during which $\sentry$ must communicate with IPs to transfer a variety of assets.  Obviously, one approach is to simply enable $\sentry$ to control the power management of the entire SoC.  However, doing so would add significant complexity to the $\sentry$ design and break the principle of using $\sentry$ as a modular subsystem.  Instead, our approach is to include a ``boot interface'' to enable communication with HOST IPs when the system bus is inactive.  Our boot interface implementation enables the communication between the compute enclave and the host processor through a generic GPIO interface, which permits interrupt-based communication with $\sentry$.

We conclude this section with an observation of the parameterized nature of the skeletal architecture and the role of security IPs.  The skeletal architecture merely provides the {\em enabling control and coordination framework} for security.  The specific protection requirements are enforced through individual security IPs and the software executed on the compute enclave.  We discuss one instance of $\sentry$ below, with specific IPs to protect against various supply-chain threats.  Obviously, the effectiveness of any $\sentry$ instance would ultimately depend both on the efficacy of the individual IP implementations to protect against a specific threat as well as the coordination of that IP with the rest of the SoCs.  The key contribution of the skeletal architecture is to provide an architectural framework for enabling such integration: any instance of $\sentry$ can simply augment the skeletal architecture with individual IPs to create a comprehensive security solution, without the need for ``re-inventing'' the coordination mechanisms for enforcing the security constraints.  With this framework, any such security strategy will be systematically broken into the following steps: 

\begin{enumerate}
    \item Identify the control and coordination functionality with the different IPs in the SoC as well as external interfaces (AMI); 
    \item Partition the communication and control functionality into software routines and hardware functionality depending on performance and overhead requirements; 
    \item Develop and integrate the hardware SCMs (if any); and 
    \item Determine wrapper functionality if the SCM strategy requires communication with other IPs.  
\end{enumerate}
\section{Instantiating $\sentry$ Skeleton For Supply-chain Confidentiality Attacks}
\label{sentrysc}
The skeletal architecture of $\sentry$  can be instantiated with specific SCMs to enable systematic protection against a variety of threat models.  In this section, we exploit this observation to create a $\sentry$ instance for protection against supply-chain confidentiality attacks.

\paragraph{Threat Model} Our threat model considers untrusted fabrication and testing facilities having access to the fabricated IC as well as rogue users after deployment.  The attacks considered include reverse engineering, counterfeiting, overproduction, and illegal recycling.   
The security requirement is to protect the design features of the SoC against such attacks.  Table~\ref{tab:threat-handled} shows the overall high-level security threats considered.  We will show detailed examples of some of the attacks under the threat model in Section \ref{usecase}.  Note that the collateral and capabilities available to the different rogue players at different stages of the chip lifecycle, {\itshape e.g.}, a fabrication/testing facility where the chip is booted for the first time may have access to the unregistered chip that can be used suppressed from the market and used as a counterfeit while at deployment in-field, the user may only have access to a chip registered with the AMI.    

\begin{remark}
The threat model for this $\sentry$ instance discussed here is inspired by the DARPA program for assurance of secure silicon (AISS) \cite{aiss}.  The threats correspondingly include supply-chain confidentiality, \eg, reverse-engineering and counterfeiting. However, note that the $\sentry$ architecture itself is agnostic to the specifics of the threat model and consequently permits protection against a diverse class of adversaries through integration of appropriate SCMs.  For instance, if the threat classes are extended to include malicious implants, integrity attacks, or fault injection attacks, the corresponding $\sentry$ instantiation would ``merely'' require an SCM for detecting/mitigating Trojan actions or fault injection incorporated as a security IP. Indeed, in previous work, Trojan detection was considered through a similar centralized subsystem by integrating fine-grained security policies \cite{tifs17}.
\end{remark}

Given the above threat model, our high-level security strategy is to use an unclonable physical signature of the chip as a chip ID to protect against IC counterfeiting by an untrusted foundry or testing facility.  Correspondingly, hardware design obfuscation is used to address reverse-engineering attacks. In our implementation, we use the unique signatures obtained from MeLPUF and the obfuscated designs using the ProtectIP \cite{protectip} framework. However, when these methods are applied in an SoC, there are architectural issues that need to be addressed.  Following are some of the representative issues.\footnote{Another challenge is to protect the sensitive communication of a newly fabricated IC, possibly in an untrusted foundry or testing facility with AMI.  Such communication is required at first power-on, {\itshape e.g.}, to enroll with AMI the chip ID collected from the PUF signature.  For the $\sentry$ instance discussed here, we assume that these communications are performed through the mediation of a trusted hardware security module (HSM). This is consistent with current industry standards \cite{IaaS}\cite{hsmchannel}.}

\begin{figure}
		\begin{center}
			\centering
			\includegraphics[width=0.7\columnwidth]{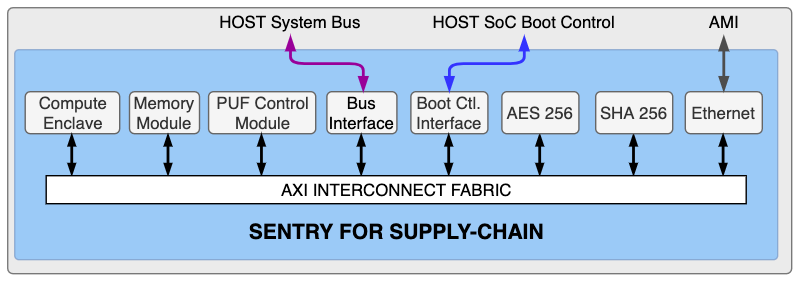}
		\end{center}
		\caption{An instantiation of $\sentry$ skeletal architecture for supply chain confidentiality protection.}
		\label{fig:sentry}
\end{figure} 

\begin{itemize}
\item How will the PUF signatures be generated and communicated?
\item How will the design obfuscation keys be provisioned by $\sentry$?  How can we make sure they are inaccessible to untrusted players?
\item How will the coordination among IPs be implemented for creating ChipID through PUF signatures?  How will the signatures be extracted from individual IPs?
\end{itemize}

$\sentry$ enables the architect to systematically analyze the questions above and develop disciplined architectural solutions to address them.

\subsection{Hardware Instantiation of the $\sentry$ Skeleton Architecture}
\begin{figure}
		\begin{center}
			\centering
			\includegraphics[width=0.7\columnwidth]{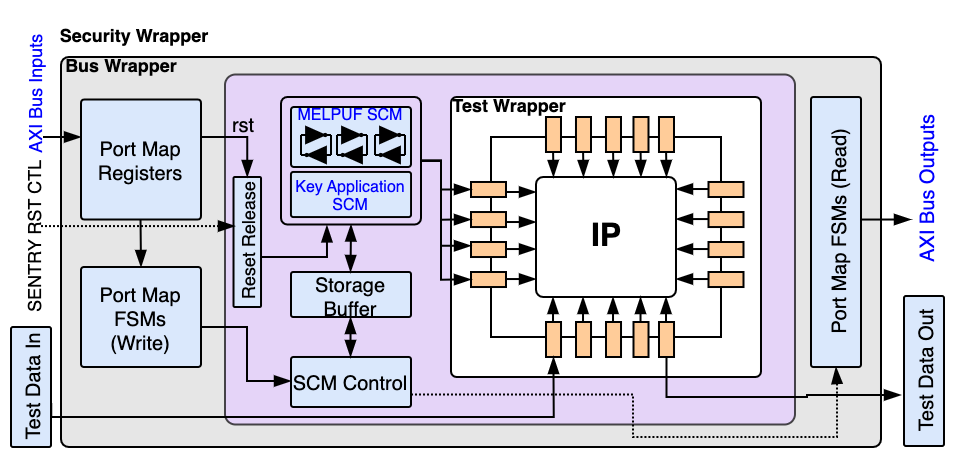}
		\end{center}
		\caption{A security wrapper (augmented generic wrapper) with integrated MeLPUF and key application SCM on a HOST IP.}
		\label{fig:wrapper}
\end{figure}

\paragraph{Security Countermeasures (SCMs)} We use SCMs to implement (1) the necessary secure storage ({\itshape e.g.}, for obfuscation keys and chip ID created from MeLPUF), (2) the PUF Control Module necessary for MeLPUF and the MeLPUF wrapper as the SCM satellite unit, and (3) the key application SCM satellite unit necessary for obfuscation, as shown in Fig. \ref{fig:wrapper}. 

\begin{remark}
SCMs (MeLPUF \cite{melpuf} and ProtectIP \cite{protectip}) are integrated with the $\sentry$ instance since they are representative technologies. Security protections in today's practice is an arms race between attack and defense, so one can envision that these technologies will be broken in the future, and likely new innovations will come in protection mechanisms superseding these technologies.  If these technologies are broken, obviously the security provided by the $\sentry$ instance incorporating them will be broken too.  However, $\sentry$ itself is an architectural framework independent of the specifics of the protection technology.  We do not take a position specifically defending these technologies.  Rather, the goal of the architecture is to enable disciplined integration of {\em diverse} protection mechanisms in SoC without requiring hand-crafted custom security implementation. 
\end{remark}

\paragraph{Enclave Software} We implement the compute enclave using the Bluespec RISC-V processor core.  The software running on the compute enclave is responsible for control and coordination of SCMS (MeLPUF and design obfuscation, a.k.a the PUF and logic locking functionality), {\itshape e.g.,} it initiates unlock requests for locked IPs during boot and initiates commands for extraction of PUF signatures from IPs during the chip enrollment process. It also mediates the coordination between $\sentry$ and the host processor during system boot, manages communication with the AMI for SoC and asset enrollment, and controls the SoC throughout its device lifecycle.

\paragraph{Security Wrappers} These are generic wrappers augmented with various two SCMs needed for enforcement of security primitives by $\sentry$. They are responsible for extracting and application of control information from individual IPs. The functionality of the wrappers are extraction of MeLPUF signatures (SCM1) and the application of obfuscation keys (SCM2) under the command of the enclave software.  Fig.~\ref{fig:wrapper} shows the wrapper design with MeLPUF and key application functionality included.  Note that the functionality is positioned amid the IEEE 1500 test wrapper and standardized AXI-4 bus wrapper.

\paragraph*{Extensibility and Modularity}
As explained above, a key aspect of $\sentry$ is its modularity and flexibility to account for various aspects of the threat model as well as the overhead constraints.  As an obvious example, the  Bluespec RISC-V core can be swapped with a custom RISC-V core to meet the design requirements. A more subtle aspect is the extensibility to handle refinement and sophistication of the threat model.  For instance, the threat model discussed above can be viewed as a ``base level'' threat model in the sense that it did not account for possible malicious access to the chip ID or obfuscation keys either from the key storage module or during communication.  Accounting for that simply requires extending the enclave software with security policies for access control \cite{7372616}, and appropriately adjusting the IPs to account for adversary capabilities, {\itshape e.g.}, if the adversary is assumed to have side channel access to stored data then the storage IP would be extended to a ``vault'', {\itshape i.e.}, tamper-proof storage that is resistant to access via side-channel \cite{cryptography6020019} \cite{9453841}.  Correspondingly, eavesdropping threats during communication can be addressed through security policies controlling encryption, which in turn extends the security IPs with crypto modules ({\itshape e.g.}, AES and SHA modules), which are already available in this instance of $\sentry$. Indeed, our current implementation, shown in Fig.~\ref{fig:sentry} accounts for an extended threat model that includes adversaries with side channel access power and the ability to eavesdrop on communications.

\begin{figure*}
		\begin{center}
			\centering
			\includegraphics[width=\textwidth]{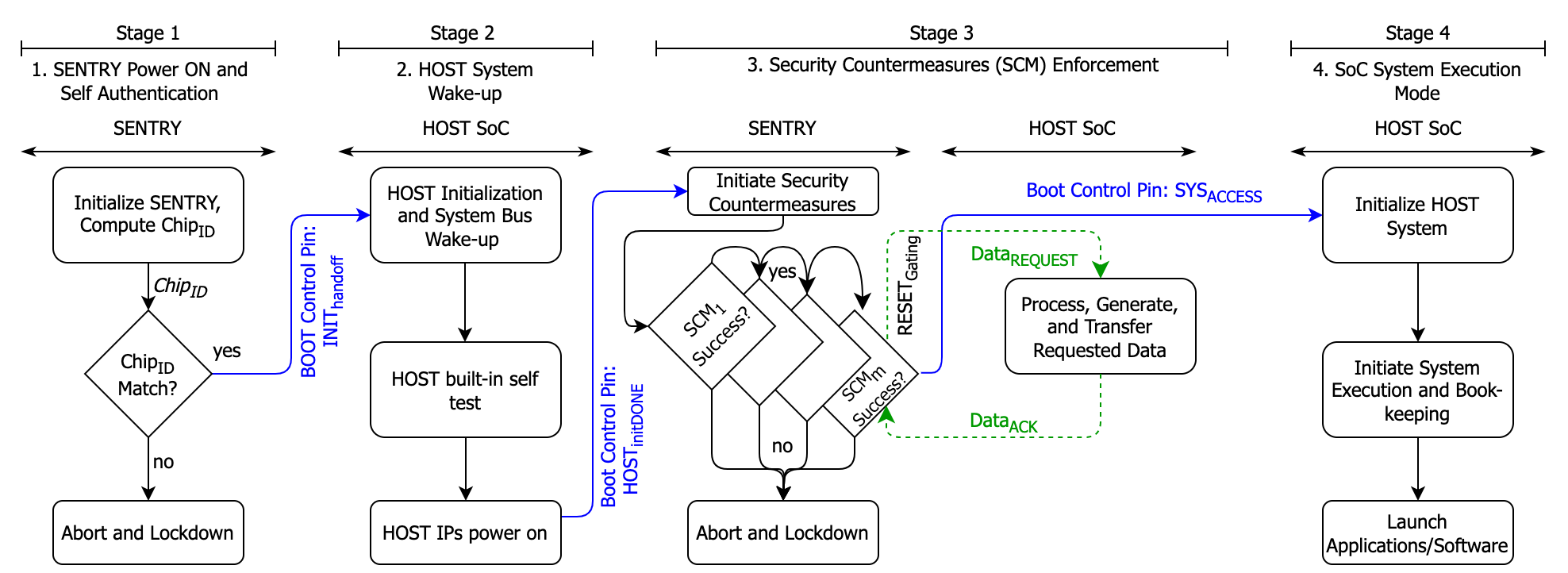}
		\end{center}
		\caption{Boot interaction flow diagram between $\sentry$ and HOST SoC via the Boot Control Interface for a Supply-chain use-case. Only communication between $\sentry$ and HOST is shown.}
		\label{fig:bootflow}
\end{figure*} 

\subsection{$\sentry$ Software Control}

The $\sentry$ skeletal architecture provides a generic hardware architecture, which enables the specific control and coordination functionality to be moved into a microcontroller firmware implementation, which is different from any boot firmware to be executed on the main processor of the SoC. The architecture provides hardware-level hooks to simplify the firmware executed in the compute enclave. Such layering results in flexible instantiations of $\sentry$, as the same skeletal architecture can then be deployed for various security mechanisms. For this instance of $\sentry$ for supply-chain, the software control consists of a 4-stage sequence between itself and the HOST SoC, as shown in Fig. \ref{fig:bootflow}. This firmware design standardizes the interaction between $\sentry$ and the HOST, which demarcates their responsibility during system boot. While $\sentry$ is responsible for \emph{only} the security aspect, the HOST is responsible for the SoC wake-up and boot. As a result, the size and complexity of the security firmware of $\sentry$ reduces significantly, eliminating the need for bootstrapping in the case of $\sentry$'s boot sequence. -- However, $\sentry$ controls of initiation of the HOST system boot after successfully booting up its security subsystem, as shown in Fig. \ref{fig:bootflow}. 

\paragraph{$\sentry$ Boot Control Sequence} The complexity of (any) boot protocol arises from three main factors: (1) self-stabilization, (2) secure management of assets and information flow, and (3) bootstrapping. In the case of supply-chain, self-stabilization includes lifecycle attestation, remote authentication with AMI, etc. Similarly, asset flow includes the transfer of PUF responses from HOST IPs to $\sentry$, application of obfuscation vectors to IPs, communication with the AMI, etc. As highlighted above, bootstrapping does not apply to this instance of $\sentry$ for supply-chain. In particular, $\sentry$ provides architectural features to close this gap and simplify the overall boot process by decoupling the actions of the host CPU to wake up different system components from the security-related actions, such as unlocking and authentication of the constituent IPs. A key requirement in enforcing SCMs at boot is to enable a secure transfer of data (assets including PUF signatures, watermarks, cryptographic keys, etc.) from the HOST to $\sentry$. An interesting conundrum for boot is the retrieval of data from the HOST to $\sentry$. Since the main system bus is inactive in the first phase of boot, $\sentry$ issues an interrupt using the \emph{Boot Control Interface} to the HOST for initialization and system bus wake-up. This is followed by an acknowledgment from the HOST to $\sentry$ indicating successful Phase 2 operations. At this stage of the boot, the system bus is active and SCMs can be enforced. SCMs can be executed in chaining or independent of preceding SCMs, contingent upon the nature of the SCMs or the devised mitigation strategy. To facilitate data retrieval from the HOST for SCM execution, each HOST IP can be triggered in isolation to retrieve the data required, while other IPs are held at reset. This protection also ensures that the data in transmission is only visible to the source IP and $\sentry$. SCM executions can be configured through firmware or security policies, allowing for modularity and extensibility in the software stack as well. After all successful SCM executions, $\sentry$ relinquishes system access to HOST via an interrupt, after which the SoC can initiate system operation and launch applications.

\begin{remark}
Note that in this instance of $\sentry$ for supply-chain, the boot control sequence accounts for hardware-based attestation, while the responsibility for system firmware authentication is delegated to the HOST processor. This is solely due to the definition of threats laid out by the threat model under consideration. It is certainly possible to migrate this functionality to $\sentry$, as has been demonstrated in other academic work \cite{riscv-boot}.  
\end{remark}

Some interesting subtleties arise in the boot control sequence for an untrusted supply-chain environment when accounting for the different device lifecycles (Fig. \ref{fig:lifecycle}). As the device transitions from one lifecycle to the next, the execution environment (including threats) and ownership of the SoC change, and consequently, the $\sentry$'s boot control needs to account for such subtle alterations. For example, after registration of the device with the AMI at the testing lifecycle, subsequent boot stages in the OEM lifecycle no longer involve device registration; instead, they account for device attestation and verification of its registration status, as shown in Fig. \ref{fig:secure_boot}. As such, some activities that need to be accounted for include:

\begin{figure*}
		\begin{center}
			\centering
			\includegraphics[width=\textwidth]{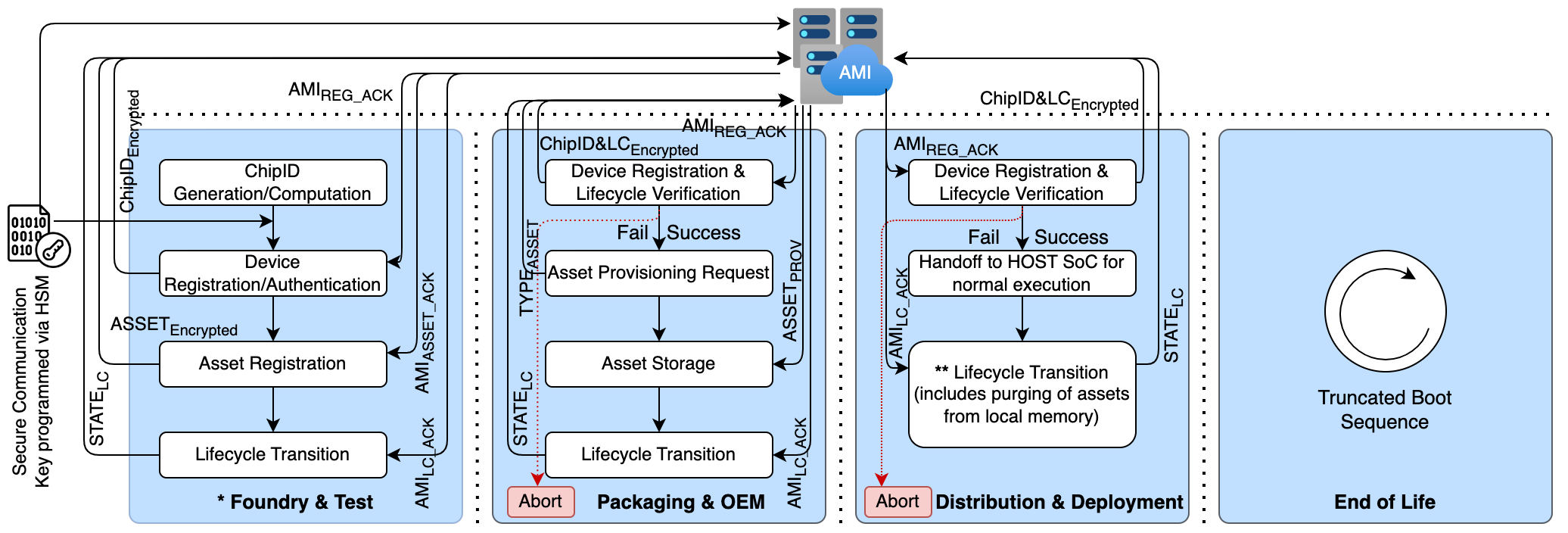}
		\end{center}
		\caption{Lifecycle management using $\sentry$'s boot control sequence. Only communication with the AMI is shown.\\
        * The SoC is under the control of the Hardware Security Module (HSM) during chip birth and registration only. \\
        ** It should be noted that lifecycle transitions from deployment to recall are under the control of the OEM. Hence, it is crucial to purge any end-user assets, firmware, and application code from the device and update the corresponding status of the device with the AMI to avoid unauthorized asset requests from the AMI.}
		\label{fig:secure_boot}
\end{figure*} 

\begin{itemize}
    \item Registration (and attestation) of system hardware integrity.
    \item Secure provisioning of identities (IDs) from AMI.
    \item Mutual authentication of components within a system or a package.
    \item Supply chain tracking, including device disenrollment.
\end{itemize}

\subsection{Device Lifecycle Support}

\begin{figure}[h]
    \begin{center}
        \centering
        \includegraphics[width=.9\textwidth]{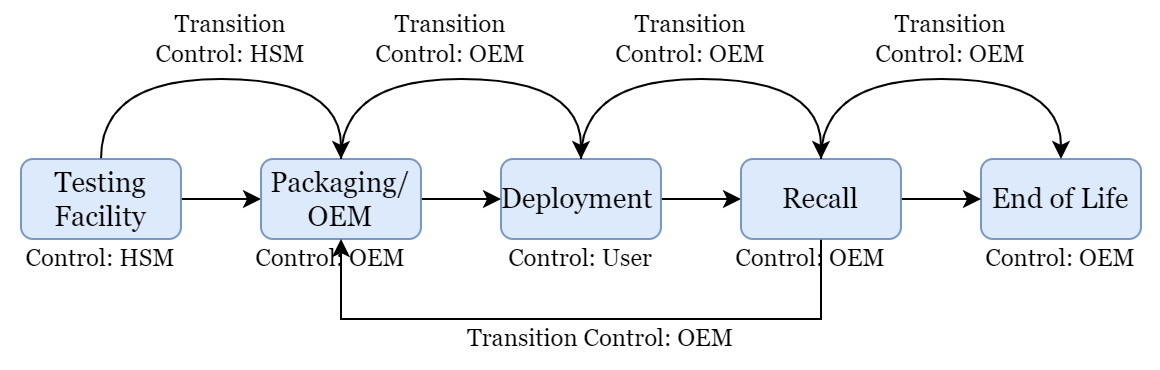}
    \end{center}
    \caption{SoC Lifecycle Stages Supported by $\sentry$. \textit{Control} represents the entity having physical access and control of the chip in the respective lifecycle and \textit{Transition Control} indicates the entity authorized to transition the lifecycle state of the device from the current state to the next.}
    \label{fig:sentry-lc}
\end{figure} 

Any microelectronics device transitions through a sequence of lifecycles.  By a ``lifecycle'', we mean the duration of time when the device is in the custody of a specific stakeholder, \eg, foundry, Original Equipment Manufacturer (OEM), user, etc.
Obviously, transitioning from one lifecycle to another impacts the targeted security profile, \eg, the degree of access of an OEM is different from that of the user. $\sentry$ provides explicit architectural support for lifecycle transitions, including the following activities.


\begin{itemize}
    \item Secure provisioning of identities (IDs) and/or firmware.
    \item Attestation of system hardware authenticity and integrity.
    \item Update of IDs.
    \item Mutual authentication of components within a system or a package.
    \item Supply chain tracking including device dis-enrollment, and optionally re-enrollment.
\end{itemize}

Fig. \ref{fig:sentry-lc} shows lifecycles supported by $\sentry$.   The idea is to consider a chip being ``born'' (powered on for the first time) in a possibly untrusted testing facility and then sequence through ``Packaging'', ``Deployment'', ``Recall'', and ``End of Life''.   As with any other architectural features, the sequence can be configured: one lifecycle stage can be split into multiple more fine-grained stages (\eg, the ``Deployment'' stage can be split to comprehend the security operations when the device passes from the custody of one user to the next), and two stages can be fused (\eg, ``Recall'' and ``End of Life'' can be fused if the device is disallowed to pass back into the open market after recall).  Nevertheless, these five stages are representative of real-life microelectronic lifecycles.   Each lifecycle has its own set of security and non-security operations, data (including assets) generation, movement, and processing as discussed below.

\smallskip \noindent
\textit{Manufacture and Test:} This lifecycle represents the steps between fabrication to acceptance when chips generate their first set of unique identifiers (ChipID). During this lifecycle, the chip is under the custody of a possibly untrusted testing facitlity, and manufacturing tests are done to the design.\footnote{These tests are generally not functional tests and involve verifying basic read/write and register access tests, stuck-at faults, X propagation, power-on tests, etc. to verify fabrication correctness to an extent.

A key activity on the first power on is the generation and registration of ChipID.\footnote{This interaction with the AMI is the only compulsory instance of communication between $\sentry$ and the AMI. All other instances of communication are not mandated.}. After successful registration of the ChipID, other assets such as the lifecycle state, PUF CRPs, obfuscation vectors, communication keys, and other relevant assets are registered or provisioned from the AMI. This process is usually done using an HSM and the ATE (Automated Testing Equipment) using which some of the security-critical assets are registered and provisioned.  ChipID and other assets in this lifecycle require the assistance of an HSM to mediate the interaction between the AMI and $\sentry$. Upon successful completion of the above-mentioned processes, the HSM transitions the lifecycle state from \texttt{FABRICATION/TEST} to \texttt{PACKAGING/OEM}.}

\smallskip \noindent
\textit{Packaging and OEM:} SoCs are then integrated after fabrication/test to a system board or system-in-package. This includes the ability to store additional assets (firmware signature, scan keys, etc.). Furthermore, a security process for mutual authentication between the SoC and the system integrator is also invoked in this lifecycle as a part of $\sentry$'s security policies.

\smallskip \noindent
\textit{Deployment:} This involves the distribution of packaged systems to end users. $\sentry$ performs self-authentication at power on at first boot in this lifecycle and upon successful authentication, releases the HOST processor to enter normal execution mode.

\smallskip \noindent
\textit{Recall:} The core idea is to ensure that the only long-term end-user changes to the chip or device are (1) physical changes due to use and (2) manufacturer or integrator-approved upgrades. At this stage, the OEM can decide on whether to re-enroll the device or decommission the device entirely. Some use cases may allow a device to re-enter the supply chain after dis-enrollment, for example through authorized device recycling or refurbishment channels. In these cases, steps for the \texttt{OEM} lifecycle for re-enrollment are repeated. After a device is dis-enrolled, it could also be disposed of. $\sentry$ ensures that all data, whether it is from the manufacturer, integrator, or OEM is removed from the device, and the AMI is also updated to reflect this new state of dis-enrollment. It is important to note that this transition might be skipped entirely. For example, a laptop could be run over by a bus which means it is effectively disposed of despite not going through the transitions.

\smallskip \noindent
\textit{End-of-Life:} In this lifecycle, no operations are allowed to happen, and the SoC enters a truncated boot sequence. Once the OEM transitions the device lifecycle to \texttt{End-of-Life}, all assets, including the ChipID, communication keys, UUIDs, etc. are erased from $\sentry$, except the lifecycle state of the device. Correspondingly, the status of the chip is also updated with the AMI as soon as the device transitions to this lifecycle, and this happens at \texttt{Recall}.

\begin{remark}
Direct communication with the AMI is required at the first boot after chip-birth during the Tesing lifecycle, as shown in Fig \ref{fig:secure_boot}. During this lifecycle, $\sentry$ provisions an asset called ``lifecycle validation key", which is used to authenticate the AMI and entities attempting to transition the lifecycle of the device. This key is provided by the OEM to the HSM. Lifecycle transition is a security functionality as different lifecycles involve distinct functions (in the form of security policies, which include authentication checks, IP unlocking, lifecycle transitions, etc.) and operations in the SoC, hence requiring a gatekeeper to authenticate such transitions. Each lifecycle transition has a unique validation key, which needs to be provided while attempting to make such transitions. The instance of $\sentry$ for supply chain in this paper uses 256-bit lifecycle validation keys for each lifecycle.
\end{remark}



\section{Supply-Chain Security Case-studies With $\sentry$}
\label{usecase}

$\sentry$ has been used to develop security solutions corresponding to a variety of supply chain threats. In this section, we discuss several representative use cases.  The case studies shown in this section are representative, but not comprehensive.  We discuss them in detail to provide a flavor of different threat mitigation designs using $\sentry$. 

\subsection{Threat Class I: SoC/IP Counterfeiting and Overproduction at Chip Birth}
\subsubsection*{Threat Description}
Consider a (possibly untrusted) fabrication and testing facility where the chip is powered on for the first time.  The threat entails the untrusted testing facility snooping ChipID during the registration, and using the snooped value to register a counterfeit chip. At this stage, the chip is born and has not yet been registered with a central facility, and the ChipID is a critical asset that needs to be safeguarded. 

\paragraph*{Security Strategy} The security mitigation makes use of the available PUF functionality in $\sentry$. The security architecture integrates PUF triggers on a (user-defined) subset of N IPs in the SoC, whose signatures are used to generate a unique and unclonable ChipID, which is registered with the AMI. Finally, to avoid the possibility of ChipID and other IDs being snooped by other IPs while they are being transferred over to $\sentry$ via the system bus, the security wrappers use the reset gating logic controlled and orchestrated by $\sentry$. Note that the HSM is in control of the SoC throughout the security enforcement process in this scenario.
\begin{figure}[h]
		\begin{center}
			\centering
			\includegraphics[width=\textwidth]{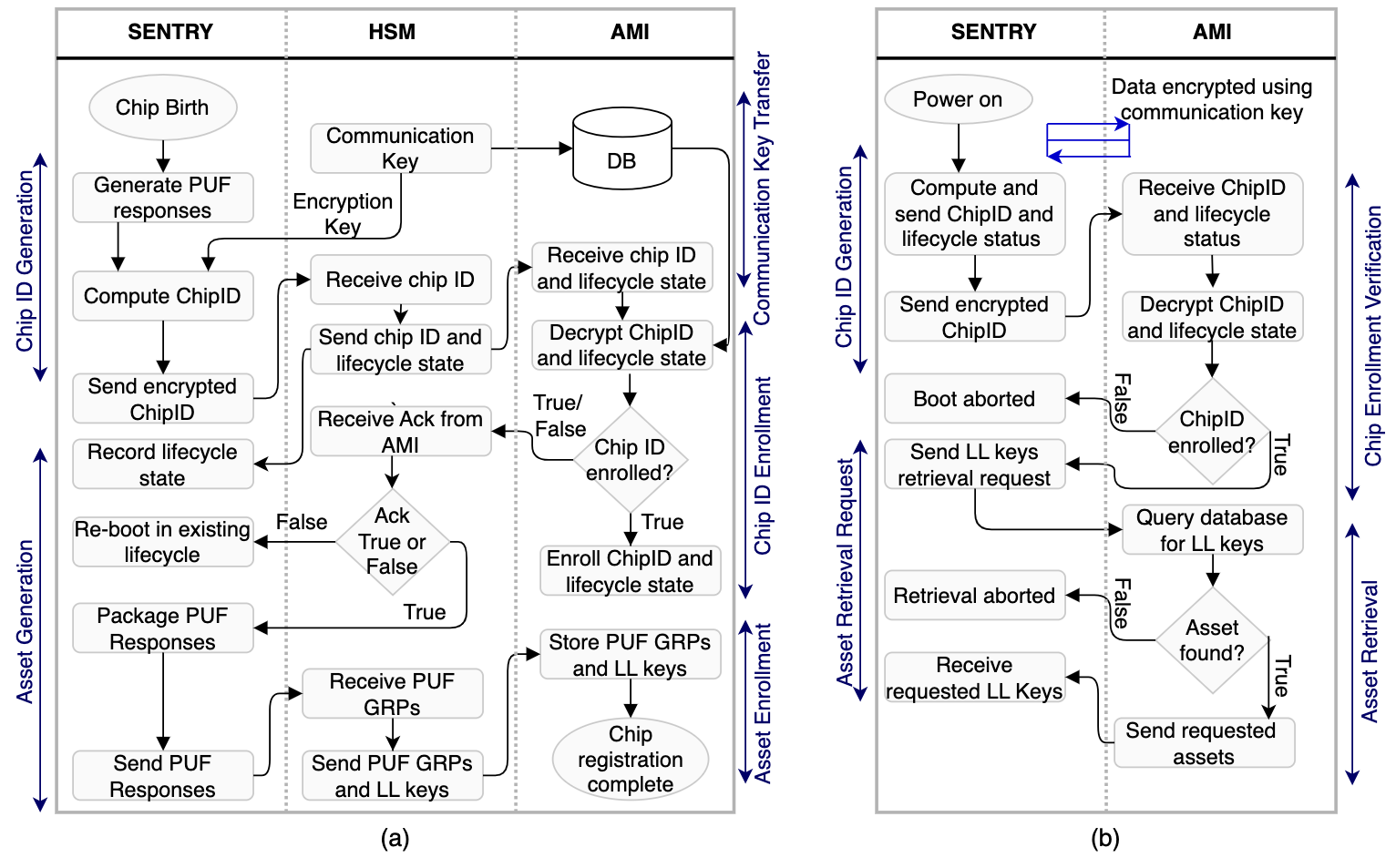}
		\end{center}
		\caption{(a) ChipID registration and asset enrollment flow (b) AMI authentication \& asset provisioning flow}
		\label{fig:usecase}
\end{figure} 
\paragraph*{Implementation}

Fig. \ref{fig:usecase}(a) shows the flow of operations for implementing the strategy above.  It is divided into three phases outlined below. 
\begin{itemize}
    \item Chip ID Computation: During the chip birth boot-up of the chip, the chip establishes a connection with the AMI using a Hardware Security Module (HSM). The chip generates PUF golden response pairs from N IPs and computes the ChipID, which, along with the lifecycle state is programmed onto the chip via the HSM. The ChipID along with the lifecycle state is encrypted using the secure communication key provided by the HSM and transferred to the AMI. 
    \item AMI Registration Check and Enrollment: If AMI realizes that the chip is already registered, the process is terminated, and a response is sent back to $\sentry$ stating that the registration was unsuccessful. If the chip operates in an illegal lifecycle, it re-boots in the previous lifecycle. If the ChipID registration is successful, the lifecycle record of the chip is updated in the AMI ledger, indicating that the chip has transitioned from the initial registration process, and subsequent registration requests from this chip will not be served.
    \item Asset Enrollment: After completion of the chip registration, $\sentry$ proceeds to register IPIDs (hashed PUF responses) and the obfuscation vectors (using the HSM) with the AMI. During the entire sequence of operations, the chip is to be physically placed inside the HSM. Assets and data transferred from $\sentry$ to the AMI are encrypted with a ``secure communication key", which is generated by the HSM and shared between both parties (chip and the AMI).
\end{itemize}

\begin{remark}
The rationale behind using the HSM in an untrusted environment (foundry and test in this case) relies on the notion that the HSM facilitates a small fraction of communication with the AMI which can be secured independently. This has been a topic of active and significant research \cite{hsmchannel}\cite{hsm-study}\cite{IaaS}. They establish how an HSM can work securely in an adversarial environment, where the key idea is modeling the HSM like an ATM, where while the HSM can be potentially physically destroyed, its actions and communication cannot be corrupted. $\sentry$ leverages this as a baseline assumption for securing communication with the AMI at chip-birth and receives a \texttt{communication key} from the HSM, using which any and all data or assets are communicated off-chip throughout the remainder of the device's operational life.

To ensure the integrity of the assets being transferred off-chip during communication with the AMI, $\sentry$ protects them using encryption and hash functions using the AES-256 and SHA-256 IPs inside its secure boundary. The ChipID is computed by XORing all values of PUF responses from all MeLPUF instances in the HOST SoC and hashing them to get a 256-bit unique ID (ChipID = \( f_{\text{SHA-256}}(PUF_{0}\oplus PUF_{1}\oplus...PUF_{i}\oplus)\)). $\sentry$ also stores the individual PUF responses and the obfuscation vectors using the same function. During communication with the AMI, each asset is encrypted using the \texttt{communication key}, using the function  \( OA = f_{\text{AES-256}}(Asset)\), where \( OA\) is any outgoing asset to the AMI.
\end{remark}

\subsection{Threat Class II: Reverse Engineering}
\subsubsection*{Threat Description} Consider a threat where an untrusted foundry, testing, or decommissioning facility (all possibly untrusted) attempts to reverse engineer the SoC (or parts of it) to extract privileged information regarding the design, IP secrets, and make alterations to the system functionality. 

\paragraph*{Security Strategy} The mitigation strategy involves the design obfuscation of unsanitized zones in the SoC. Unsanitized zones are zones that need to be protected against reverse engineering attacks. This is done using the ProtectIP obfuscation scheme \cite{protectip} technology by obfuscating parts of the design (unsanitized zones).

\begin{figure}
	\begin{center}
		\centering
		\includegraphics[width=0.7\columnwidth]{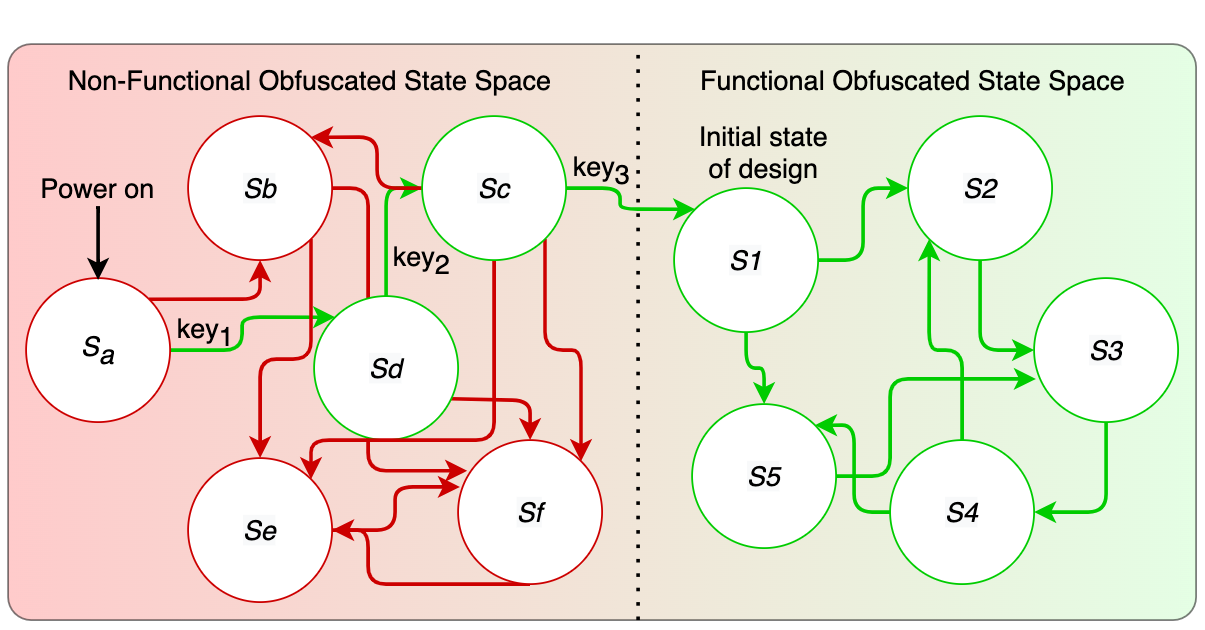}
	\end{center}
	\caption{A modified state-transition diagram for an obfuscated design with additional transition states.}
	\label{fig:obf}
\end{figure} 

\paragraph*{Implementation} $\sentry$ uses an obfuscated netlist of the IP, whose design is functionally equivalent to the original design. Fig. \ref{fig:obf} shows the modified netlist of the design where only one state path unlocks the design into its functional obfuscated space through the application of a series of unlocking keys. In this implementation, N IPs in the SoC are protected against reverse-engineering attacks. The unlocking procedure is divided into two phases outlined below. 
\begin{itemize}
    \item Asset Provisioning: The logic locking keys for the obfuscated IPs are provisioned by $\sentry$ upon successful registration \& authentication of the chip with the AMI post-testing, as shown in Fig. \ref{fig:usecase}(b). Each key is stored on the secure on-chip memory inside $\sentry$ with a unique identifier indicating the corresponding IP. 
    \item Key Application for Unlocking: The logic locking keys are transferred to the security wrapper of obfuscated IP instances during system boot and it programs the keys/activation package, traversing through the obfuscated FSM 
    \begin{math}
      (Sa \xrightarrow{} Sd \xrightarrow{} Sc \xrightarrow{} S1), 
    \end{math}
    as shown in Fig. \ref{fig:obf}, and the IP is unlocked. Obfuscated regions can similarly be locked back to their original state by $\sentry$ if specific violations are detected or security policies are violated.
\end{itemize}

\subsection{Threat Class III: Illegal Recycling at an Untrusted Facility}
\textit{Threat Description:} Consider a case where a chip is sent to a recycling facility to be decommissioned and recycled for components or materials. The facility (possibly untrusted) may attempt to re-brand the chip with new markings and physical identification and sell it in the open market for financial gains. While preventing physical alterations to the chip is beyond the scope, the management and control over the activation and enrollment of decommissioned chips present a realistic approach to protect against such attacks.

\smallskip

\textit{Security Strategy:} 
$\sentry$ maintains the chip's registration and lifecycle status with the AMI using which the chip can be tracked throughout the lifespan of the chip and control re-enrollment back into the supply chain. The lifecycle status is modifiable under specific circumstances by \textit{authorized} entities (see Fig. \ref{fig:sentry-lc}), and once a chip transitions to the decommissioning lifecycle, the AMI marks the chip as \textit{deactivated} or \textit{decommissioned}, upon which, it cannot be re-enrolled into the supply chain. $\sentry$ also maintains its current lifecycle value locally using which it boots on power-on. Lifecycle verification is the first check done during the first boot at power-on in the respective lifecycles. In the case where it does, the lifecycle status and the AMI record ensure that the chip falls into a truncated boot sequence, rendering it inoperable.

\smallskip

\textit{Implementation:} 
The AMI keeps a record of the chip metadata to track the illegal recycling of the chip. When the chip transitions from OEM to deployment or decommissioning, the metadata (ChipID, current lifecycle) of the chip is updated accordingly in the AMI ledger.  Upon first power on in each lifecycle, the chip undergoes a truncated boot sequence to validate the existing lifecycle after successful validation of ChipID. This sequence of steps is performed on every chip lifecycle transition before the chip is booted up in the new lifecycle to ensure the boot-up, hand-swap or further actions are legal. If the validation process fails, $\sentry$ boots the chip in the preceding lifecycle.  Now, if a malicious recycling facility repackages the chip and attempts to sell it in the open market, the chip will not be able to successfully boot up as the current lifecycle status of the chip indicates the chip has transitioned to decommissioning.
\section{Overhead Evaluation of $\sentry$ Instance for Supply-chain}
\label{experiment}

Integrating a security engine solution involves overhead in various parameters, including power, area overheads, and timing delays. In this section, we perform an empirical evaluation of the overhead of the $\sentry$ instance for supply-chain security. To our knowledge, there is no other comprehensive SoC security architecture for supply-chain attacks for systematically implementing ad-hoc supply-chain security solutions to enable direct comparison with $\sentry$.  However, our evaluation of $\sentry$ demonstrates that it has a minimal area footprint and incurs a very small cost in all the relevant overhead parameters.

\begin{figure}[h]
\begin{center}
\begin{tabular}{cc}
\begin{minipage}{0.49\textwidth}
    \includegraphics[width=\columnwidth]{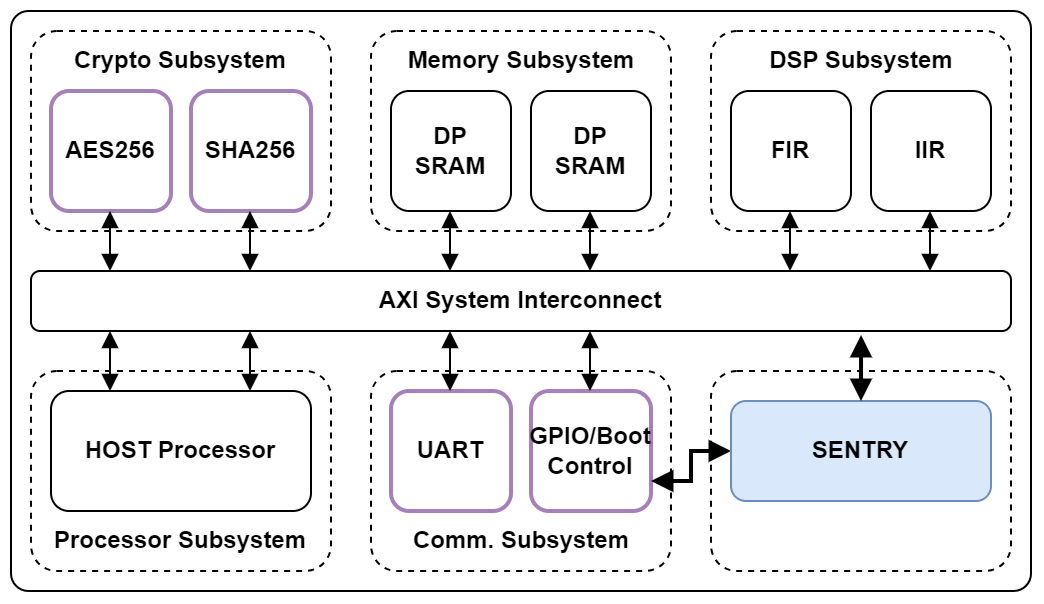}
\end{minipage}
&
\begin{minipage}{0.49\textwidth}
 \includegraphics[width=\columnwidth]{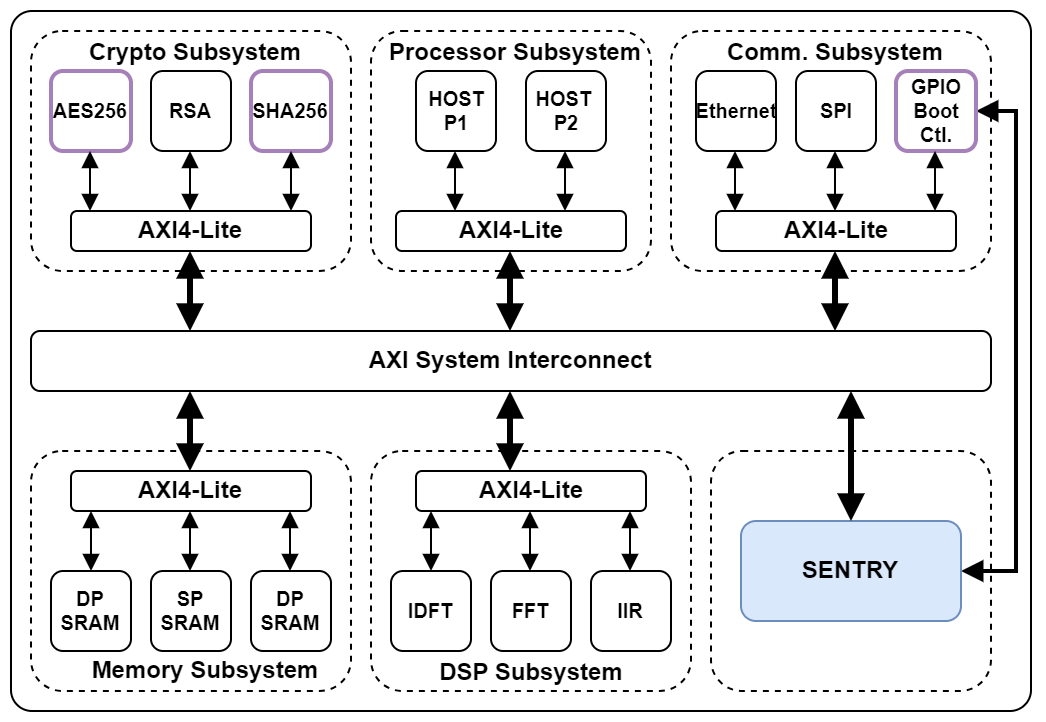}
\end{minipage}
\\ 
(a) & (b) \\
\end{tabular}
\end{center}
\caption{SoC models used for $\sentry$ evaluation.  (a) Single-bus SoC: This SoC model has a single shared-bus architecture.  (b) Multi-bus SoC: This SoC model has an AXI4 interconnect as the primary bus and five AXI4-Lite buses as subsystem interconnect. Five AXI4-Lite bus-based peripherals/hardware accelerator subsystems are connected to the AXI4 bus through standard bus bridges. Both designs have $\sentry$ integrated.}
\label{fig:socs}
\end{figure}

To address the dearth of available open-source SoCs for evaluation of $\sentry$, we developed two  SoCs, Single-bus and Multi-bus SoC architectures inspired by industry designs shown in Fig. \ref{fig:socs} and integrated $\sentry$ into the SoC testbeds.\footnote{The SoCs are developed through an automated CAD flow \cite{soccom}. We have extended the flow to generate $\sentry$ instances for security. The flow can currently generate bus-based SoCs and supports single clock and reset domains. Note that $\sentry$ instance is a standalone subsystem, similar to IPs in the SoC, thus simplifying its integration process into SoCs like any IP block, provided the interface requirements for communication with the HOST processor are satisfied.}

The Single-bus SoC is an area and power-efficient implementation of an SoC for IoT devices targeting low-power applications. The Multi-bus SoC is significantly more complex than Single-bus and represents a tiled-hierarchical architecture with application-specific subsystems. Both SoCs are functional SoC implementations developed to verify the efficacy and cost evaluation of $\sentry$. $\sentry$ was also integrated with the MIT CEP SoC \cite{mitcep}. All these designs were synthesized on two technology nodes (LEDA 250nm and GSCL 45nm), and the overhead ratios of SENTRY with the mentioned SoC architectures are provided in Table \ref{tab:soc_overhead}. Standalone $\sentry$ area and power metrics are provided in Table \ref{tab:asic}.

The augmentation includes two layers of security primitives: (1) design obfuscation using the ProtectIP obfuscation scheme, and (2) 256-bit MeLPUF instances have also been inserted for the construction of ChipID. These instances are obfuscated with a 512-bit vector. For unlocking, these vectors are sequentially applied over \emph{P} clock cycles. Each keyframe is fragmented based on the width of the input signal width for the underlying IP (excluding clock and reset), and the number of iterations is calculated.

\subsection{$\sentry$ Area Overhead}
We synthesized SoCs with $\sentry$ integrated onto two ASIC libraries, (1) LEDA (250nm) and (2) GSCL (45nm) using Synopsys Design Compiler T-2022.03-SP5, and a comparative analysis is provided in Table \ref{tab:soc_overhead}. The major contributing components to area and power overheads were the Bluespec RISC-V core, AXI interconnect, and the 64KB system memory. The security wrappers augmented onto the IPs in the SoC in Fig. \ref{fig:socs} (indicated with the purple outline) also contribute to the area overheads of $\sentry$ although they are beyond the $\sentry$ boundary. This is because the security primitives implemented by $\sentry$ depend on the security wrappers to enforce those security primitives. The overheads from the 256-bit MeLPUF instances in the testbed SoC are also counted because they are used to generate the ChipID of the SoC. This overhead estimation of the ChipID asset is done by evaluating the number of gates required for each bit of the asset.  As such, the area overhead in terms of NAND-2 equivalent gates can be calculated using the expression:
\begin{equation}
Overhead_{PUF} = \sum_{0}^{N}\left\{ChipID_{bits}  \times 6\right\}
\label{eqn:chipIDequation}
\end{equation}

The overall overhead of $\sentry$ with all its distributed SCM controls can be calculated using the expression:
\begin{equation}
Overhead_{Total} = Overhead_{SENTRY} + \sum_{0}^{N}Sec.\ Wrapper_{um^{2}} + Overhead_{PUF}
\label{eqn:sentryoverhead}
\end{equation}
For a standardized $\sentry$ architecture for supply-chain, we concluded that such overheads (\ref{eqn:sentryoverhead}) in terms of both power and area would be minimal, as shown in Table \ref{tab:asic}.

\begin{table}[]
\centering
\def\arraystretch{1.1}
\caption{Area \& Power Metrics of $\sentry$}
\label{tab:asic}
\begin{tabular}{lccc}
\hline
\multicolumn{1}{c}{\textbf{Technology}} & \textbf{Die Area (um2)} & \textbf{Dyn. Power (mW)} & \textbf{Leak. Power (mW)} \\ \hline
\begin{tabular}[c]{@{}l@{}}LEDA \\ 250 nm\end{tabular}  & 12783906 & 551  & 15  \\ \hline
\begin{tabular}[c]{@{}l@{}}GSCL45\\ 45 nm\end{tabular}  & 756716   & 84.2 & 3.5 \\ \hline
\end{tabular}
\end{table}
\begin{table}[htbp]
\centering
\caption{Area \& Power Overhead of $\sentry$ integrated with SoC Testbeds}
\def\arraystretch{1.3}
{
\begin{tabular}{lllllll}
\toprule
\textbf{SoC} & \multicolumn{2}{c}{\textbf{Baseline Statistics}} & \multicolumn{2}{c}{\textbf{with SENTRY}} & \multicolumn{2}{c}{\textbf{Overheads}} \\
\cmidrule{2-7}          & \multicolumn{1}{p{5.865em}}{\textbf{Area (um2) LEDA}} & \multicolumn{1}{p{5.725em}}{\textbf{Area (um2) GSCL}} & \multicolumn{1}{p{5.775em}}{\textbf{Area (um2) LEDA}} & \multicolumn{1}{p{5.82em}}{\textbf{Area (um2) GSCL}} & \multicolumn{1}{p{4.045em}}{\textbf{Average Area (\%)}} & \multicolumn{1}{p{4.68em}}{\textbf{Average Power (\%)}} \\ \hline
Single-SoC & 82281134      & 3943861     & 95065040      & 4700577     & 17.50     & 17.89 \\ \hline
Multi-SoC  & 106245988     & 4561205     & 119029894     & 5317921     & 14.00     & 12.45 \\ \hline
MIT CEP    & 141211642     & 5887352     & 153995548     & 6644068     & 10.50     & 10.91 \\
\bottomrule
\end{tabular}%
}
\label{tab:soc_overhead}%
\end{table}%

\begin{table}[]
\centering
\caption{Area \& Power Overhead of Security Wrapper Instances From Synthesis on GlobalFoundries GL12LPP (12nm)}
\label{tab:wrapper_overhead}
\def\arraystretch{1.3}
\begin{tabular}{llllllll}
\toprule
\multirow{2}{*}{\textbf{IP}} & \multicolumn{2}{c}{\textbf{Baseline Statistics}}                                       & \multicolumn{2}{c}{\textbf{with Security Wrapper}}                                                                                & \multicolumn{3}{c}{\textbf{Overheads}}                                                                                \\ \cline{2-8} 
 & \textbf{Area (um2)} & \textbf{\begin{tabular}[c]{@{}l@{}}Gate \\ Count (K)\end{tabular}} & \textbf{\begin{tabular}[c]{@{}l@{}}Area\\ (um2)\end{tabular}} & \textbf{\begin{tabular}[c]{@{}l@{}}Gate \\ Count (K)\end{tabular}} & \textbf{\begin{tabular}[c]{@{}l@{}}Area\\ (\%)\end{tabular}} & \textbf{\begin{tabular}[c]{@{}l@{}}Gate \\ Count (\%)\end{tabular}} & \textbf{\begin{tabular}[c]{@{}l@{}}Total \\ Power (\%)\end{tabular}} \\ \hline
AES-256                      & 58628.10           & 404.00                                                            & 59049.44                                                      & 406.90                                                            & 0.70                                                         & 0.70                                                                & 0.91                                                                 \\ \hline
SHA-256                      & 2691.80            & 18.55                                                             & 2800.62                                                       & 19.30                                                             & 4.04                                                         & 4.04                                                                & 1.82                                                                 \\ \hline
UART                         & 2236.30            & 15.41                                                             & 2323.36                                                       & 16.01                                                             & 3.89                                                         & 3.89                                                                & 2.46                                                                 \\ \hline
GPIO                         & 1352.50            & 9.32                                                              & 1423.2                                                        & 9.87                                                              & 5.90                                                         & 5.90                                                                & 2.85                                                                 \\
\bottomrule
\end{tabular}
\end{table}

\subsection{Security Wrapper Area Overhead}
Security primitives enforced by $\sentry$ require IPs to be augmented with the security wrapper. Since in a typical SoC, all IPs are not secured, we compute the overheads using representative IPs. To obtain the representative overhead values for the security wrappers, each wrapper instance was synthesized for the GlobalFoundries GL12LPP 12nm technology.  Table \ref{tab:wrapper_overhead} shows the results. Note that the overheads are minimal ({\itshape e.g.}, 0.70\% area overhead for AES-256) when compared to the overall footprint of the underlying IP. The security wrapper overhead remains approximately constant in terms of the gate count and only varies due to the variation in the size of the buffers and MUXs implemented to switch data based on different addresses for read and write operations, based on the memory map of the IP.

\subsection{Evaluation of Security Operations}
To determine the timing delays incurred from security operations, we analyze  (1) ChipID authentication, and (2) Obfuscated design unlocking as a function of the ChipID size and the obfuscation key vector size, respectively. Fig. \ref{fig:auth_graph} shows the variation in time taken to complete the ChipID authentication sequence with different ChipID sizes. Note that the range of widely used ChipID size ranges from 128 bits to 512 bits. In this range,  the delay incurred due to the authentication operation is negligible. Fig. \ref{fig:ip_unlock} shows the variation in time delay when the unlocking key size is increased. We analyze this delay using the UART and AES-256 instances in the HOST SoC. The difference between the UART and AES-256 for the same key sizes is from the width of input signals to each instance. In this case, the width of the input signals of AES-256 is greater than UART, thus, larger segments of keys can be applied to the IP FSM at once, resulting in reduced unlocking delay. The feasible key sizes for unlocking are estimated to be from 192 bits to 512 bits, and in this range, the delay incurred due to this operation is minimal. 
\begin{figure}
\centering
\begin{minipage}{.5\textwidth}
  \centering
  \begin{tikzpicture}
    \begin{axis}[stack plots=y,
    xmode=log,
    log basis x={2},
    xlabel=ChipID Size (bits),
    ymin = 0,
    grid=major,
    ylabel= Time to Authenticate ChipID (ps),
    legend entries={AES-256 with 256-bit MeLPUF},
    legend pos=north west,
    legend cell align=left,
    width=6.5cm,scale only axis,
    ]
    
\addplot[teal,mark=square*]
    coordinates {(128,320) (256,680) (512,1270) (1024,2540) (2048,4870)};
    \end{axis}
\end{tikzpicture}
  \caption{ChipID size vs authentication delay chart.}
  \label{fig:auth_graph}
\end{minipage}%
\begin{minipage}{.5\textwidth}
  \centering
  \begin{tikzpicture}
    \begin{axis}[stack plots=y,
    xmode=log,
    log basis x={2},
    xlabel=Logic Locking Key Size (bits),
    ylabel= Time to Unlock Locked IP (ps),
    grid=major,
    ymin = 0,
    legend entries={AES-256, UART, SHA-256, GPIO},
    legend pos=north west,
    legend cell align=left,
    width=6.5cm,scale only axis,
    ]
\addplot[violet,mark=oplus*]
    coordinates {(128,120) (256,240) (512,480) (1024,960) (2048,1920)};
\addplot[teal,mark=pentagon*]
    coordinates {(128,260) (256,520) (512,1020) (1024,2040) (2048,4080)};
\addplot[black,mark=triangle*]
    coordinates {(128,110) (256,220) (512,440) (1024,880) (2048,1760)};
\addplot[magenta,mark=square*]
    coordinates {(128,250) (256,500) (512,1000) (1024,1500) (2048,2500)};
    \end{axis}
\end{tikzpicture}
  \caption{IP unlocking key size vs unlocking delay chart.}
  \label{fig:ip_unlock}
\end{minipage}
\end{figure}
\section{Related Work}
\label{related-work}

There has been significant research on developing systematic and streamlined frameworks for security architecture and the implementation of modern SoC designs. We categorize the related work in two broad categories: \textit{(1) Centralized Platforms and Standards}, as shown in Table \ref{tab:comparison-chart}, and \textit{(2) Heterogeneous and Isolated Solutions.}

\subsection{Centralized Security Platforms and Standards}
Security frameworks have been designed to enable isolation of functionality at different levels of trustworthiness through Trusted Execution Environments (TEEs). One well-known TEE architecture is the Trusted Platform Module (TPM) \cite{tpm}. TEE frameworks like Apple Secure Enclave \cite{applese}\footnote{The Apple SE produces a modular security subsystem and is architecturally similar to $\sentry$. Apple security enclave has a Root of Trust processor (similar to the compute enclave in the case of $\sentry$), which controls a subsystem with other IPs. However, it is primarily a hardware-based key manager for data integrity and privacy \cite{applekeymanager}.}, ARM TrustZone \cite{arm}, and Intel SGX \cite{intelsgx} were developed to provide segregation of the area of memory and CPU that is protected from the rest of the CPU using encryption. Security enclaves like Apple's and similarly others can be re-targeted for supply chain security, --- however, re-purposing would include significant complexity and changes in architecture. Doing so without innate architecture support generally implies a high software and design complexity. 

RealSE \cite{realse} presents another security engine candidate, built around providing security for crypto-based processors and devices. Several other works \cite{sastry2014method} \cite{miettinen2014conxsense} have also been proposed involving custom security policies to control access privileges and roles but do not present a centralized and reusable architecture solution that can be customized for diverse SoC security challenges. Basak {\itshape et al.} \cite{6915693} proposed an SoC-level solution (IIPS) for supply-chain security threats including counterfeiting, trojan attacks, etc. However, IIPS misses out on critical components like secure asset provisioning, device lifecycle management, device enrollment, handshaking, and secure boot control with the HOST processor.  IIPS was further extended to E-IIPS \cite{7372616} proposed an architecture to provide security against system-level trojans stemming from untrusted IPs affecting other IP cores or interconnect fabrics using run-time monitoring by security policies. It does not take into account other supply chain threats discussed above and also does not present a flexible architecture like $\sentry$ to augment its instance to support other security primitives. However, the solution proposed in E-IIPS can be integrated into $\sentry$ to provide security assurance against system-level trojans, as highlighted in Section \ref{sentrysc}. Moreover, both of these lack hardware and software-level SCM configurability. Another SoC security architecture and CAD framework \cite{CAD} proposes a hardware patch and CAD flow for implementing security policies in-field on FPGAs, and lacks in providing any security against supply chain attacks like overproduction, counterfeiting, reverse engineering, illegal recycling, etc. While the idea of using security wrappers to tap low-level registers and scan for data is shared with $\sentry$, their architectural integration, coordination, and access control are possible only using hooks provided by $\sentry$. Note that each of these works primarily targets one specific threat class.  Incorporating heterogeneous security primitives on an SoC level needs to account for the entire execution timeline of the SoC, and the various entanglements that come with it. None of the above solutions deal with such entanglements.

SAP \cite{sap} is a security architecture for enabling SoC authentication and secure asset provisioning in untrusted environments and chip authentication during deployment. However, SAP misses out on protection against reverse engineering attacks, which is a key component of supply-chain threats. Moreover, the architecture proposed in SAP is not extensible to provide support for reverse engineering solutions. Another key ingredient of security is its intervention during the boot process of the device, which is not mentioned in the paper. SAP was extended to CHSM \cite{chsm}, which does provide security assurance against reverse engineering attacks along with others, targeted for chiplets and 3DICs. However, re-purposing CHSM for SoC-based architectures would require significant and complex modifications in the architecture due to the difference in the attack surface between SoCs and chiplets. Further convoluting the transformation of the architecture for SoCs is that many of the security primitives used in CHSM exploit the 3D and chiplet properties, which are not present in SoC-based architectures. The paper also does not mention the integration of security provided by CHSM with the overall boot process of the underlying device.

PROTECTS \cite{PROTECTS}, a secure protocol for provisioning security assets from an external entity (\textit{i.e.} AMI) to the SoC in an untrusted fabrication and testing facility. This is only a small facet of the architecture proposed in $\sentry$, which involves communication with the AMI at chip-birth for asset provisioning. PROTECTS, unlike $\sentry$, however, does not solve the architectural problem of heterogeneous integration of diverse security primitives required to be deployed in an SoC environment to systematically protect against supply-chain attacks. To our knowledge, $\sentry$  is the first comprehensive architecture that comprehends requirements and constraints in integrating security solutions for an SoC, supports the entire SoC lifecycle,  and accounts for control and coordination with the HOST processor, as well as protection of assets in its boundary.


\begin{table}[]
\centering
\caption{A Comparative Overview of $\sentry$ With Other Security Architectures }
\label{tab:comparison-chart}
\def\arraystretch{1.1}
\begin{tabular}{lcccccc}
\toprule
Architecture &
  \begin{tabular}[c]{@{}c@{}}Counterfeit\\ Protection\end{tabular} &
  \begin{tabular}[c]{@{}c@{}}RE\textsuperscript{\#}\\ Protection\end{tabular} &
  \begin{tabular}[c]{@{}c@{}}Lifecycle\\ Management\end{tabular} &
  \begin{tabular}[c]{@{}c@{}}Trojan\\ Detection\end{tabular} &
  \begin{tabular}[c]{@{}c@{}}Resource\\ Isolation\end{tabular} &
  \begin{tabular}[c]{@{}c@{}}Asset\\ Provisioning\end{tabular} \\ \bottomrule
\begin{tabular}[c]{@{}l@{}}Apple Secure\\ Enclave~\cite{applese}\end{tabular} &
  {\color[HTML]{CB0000} ×} &
  {\color[HTML]{CB0000} ×} &
  {\color[HTML]{009901} \checkmark} &
  {\color[HTML]{009901} \checkmark} &
  {\color[HTML]{CB0000} *} &
  {\color[HTML]{009901} \checkmark} \\ \hline
OpenTitan~\cite{opentitan} &
  {\color[HTML]{009901} \checkmark} &
  {\color[HTML]{CB0000} ×} &
  {\color[HTML]{CB0000} ×} &
  {\color[HTML]{CB0000} ×} &
  {\color[HTML]{CB0000} ×} &
  {\color[HTML]{009901} \checkmark} \\ \hline
ARM TrustZone~\cite{arm} &
  {\color[HTML]{CB0000} ×} &
  {\color[HTML]{CB0000} ×} &
  {\color[HTML]{CB0000} ×} &
  {\color[HTML]{009901} \checkmark} &
  {\color[HTML]{CB0000} ×} &
  {\color[HTML]{009901} \checkmark} \\ \hline
E-IIPS~\cite{tifs17} &
  {\color[HTML]{009901} \checkmark} &
  {\color[HTML]{CB0000} ×} &
  {\color[HTML]{CB0000} ×} &
  {\color[HTML]{009901} \checkmark} &
  {\color[HTML]{CB0000} ×} &
  {\color[HTML]{CB0000} ×} \\ \hline
SAP~\cite{sap} &
  {\color[HTML]{009901} \checkmark} &
  {\color[HTML]{CB0000} ×} &
  {\color[HTML]{CB0000} ×} &
  {\color[HTML]{CB0000} ×} &
  {\color[HTML]{CB0000} ×} &
  {\color[HTML]{009901} \checkmark} \\ \hline
CHSM (3DIC)~\cite{chsm} &
  {\color[HTML]{009901} \checkmark} &
  {\color[HTML]{009901} \checkmark} &
  {\color[HTML]{CB0000} ×} &
  {\color[HTML]{009901} \checkmark} &
  {\color[HTML]{CB0000} ×} &
  {\color[HTML]{009901} \checkmark} \\ \hline
$\sentry$ &
  {\color[HTML]{009901} \checkmark} &
  {\color[HTML]{009901} \checkmark} &
  {\color[HTML]{009901} \checkmark} &
  {\color[HTML]{009901} \checkmark} &
  {\color[HTML]{CB0000} ×} &
  {\color[HTML]{009901} \checkmark} \\ \bottomrule
\end{tabular}
 \begin{tablenotes}
      \small
      \item \# Reverse Engineering.
      \item \textcolor{red}{*} No public information available. Apple's security infrastructure is controlled by its software stack.
    \end{tablenotes}
\end{table}

\subsection{Customized Security Solutions}
There have also been customized solutions for individual supply-chain threats ({\itshape i.e.} overproduction, reverse engineering, etc).  However, these approaches do not explore the question of integration requirements and system-wide coordination of these solutions in SoCs with custom security requirements.

There has been significant research on security protocols for safeguarding IP and IC designs against overproduction and counterfeiting attacks \cite{ray2015security} \cite{ray2017system} \cite{peeters2015soc}. Such techniques include IC metering \cite{wei2011robust}, Hardware Metering \cite{koushmetering}, CDIRs (Combating Die/IC Recovery) \cite{6241582}, and PUFs \cite{10.1145/586110.586132}. Techniques for mitigating reverse engineering attacks include camouflage, design obfuscation, split fabrication \cite{7858390}, etc. Camouflage is a layout-level mitigation strategy where multiple dummy contact points are added into the layout of the chip, altering the layout representation of the chip \cite{6881480}\cite{9039593}. Design obfuscation techniques \cite{protectip} include integrating additional logic in the design \cite{5247148}, and adding futile inputs and outputs to mask the target functionality. 
SCARe \cite{8237209} is an SRAM-based technique to detect the aging of SRAM memory elements to determine the aging of SRAM cells in embedded devices. Other SRAM-based age detection techniques \cite{8741032}. Other age detection techniques include EM on-chip imaging \cite{7372562}, and approximation of PUF signature variations due to aging \cite{7495581}. 
\section{Conclusion}
\label{conclusion}

In this paper, we introduced a security paradigm $\sentry$, a skeletal security architecture for providing systematic security insertion and assurance in SoC designs for a wide variety of security use cases, and not limited to the supply chain. Architecturally, $\sentry$ is a plug-and-play subsystem composed of a microprocessor and a collection of custom IPs that can be integrated with an SoC design through a standardized set of interfaces. We outlined the methodology for deriving security architecture instances by following a threat-specific driven architecture philosophy and showcased the paradigm's viability for providing security against a wide variety of supply-chain threats. $\sentry$ has been designed from the ground up to provide resiliency against a spectrum of threats across the SoC lifecycle and enables systematic integration of a variety of security requirements without having to perturb the design functionality of the different SoC subsystems. Furthermore, $\sentry$ is built out of open-source components and exploits the RISC-V design ecosystem to achieve extensibility, configurability, and robustness.  Finally, $\sentry$ achieves this with a very small hardware footprint, making it a viable security solution for low-power SoCs.  

In future work, we will extend $\sentry$ with more use cases and threat models.  We will also explore more optimization opportunities and the feasibility (and challenges) of integrating $\sentry$ with complex industrial SoCs.

\bibliographystyle{unsrt}  
\bibliography{templateArxiv}

\end{document}